\documentclass[aip,amsmath,amssymb,reprint,superscriptaddress]{revtex4-1}
\usepackage[utf8x]{inputenc}
\usepackage{textcomp}
\usepackage{graphicx,subcaption}
\usepackage[outdir=./]{epstopdf}
\usepackage{bm}
\usepackage{caption}
\usepackage{subcaption}
\usepackage{algorithm}
\usepackage{float}
 \usepackage[noend]{algpseudocode}
  \def\BState{\State\hskip-\ALG@thistlm}

\begin{document}


\title{Causal network discovery by iterative conditioning: comparison of algorithms}

\author{Jaroslav Hlinka}
\affiliation{Institute of Computer Science, Czech Academy of Sciences, Pod vodarenskou vezi 271/2, 182 07, Prague, Czech Republic}
\affiliation{National Institute of Mental Health, Topolová 748, 250 67, Klecany, Czech Republic}

\author{Jakub Ko\v{r}enek}
\affiliation{Institute of Computer Science, Czech Academy of Sciences, Pod vodarenskou vezi 271/2, 182 07, Prague, Czech Republic}
\affiliation{Faculty of Nuclear Sciences and Physical Engineering, Czech Technical University, Břehová 7, 115 19, Prague, Czech Republic}

\date{\today}

\begin{abstract}

Estimating causal interactions in complex dynamical systems is an important problem encountered in many fields of current science.
While a theoretical solution for detecting the causal interactions has been previously formulated in the framework of prediction improvement, it generally requires the  computation of high-dimensional information functionals  -- a situation invoking the curse of dimensionality with increasing network size. Recently, several methods have been proposed to alleviate this problem, based on iterative procedures for assessment of conditional (in)dependences.
In the current work, we bring a comparison of several such prominent approaches. This is done both by theoretical comparison of the algorithms using a formulation in a common framework, and by numerical simulations including realistic complex coupling patterns.
The theoretical analysis highlights the key similarities and differences between the algorithms, hinting on their comparative strengths and weaknesses. The method assumptions and specific properties such as false positive control and order-dependence are discussed. 
Numerical simulations suggest that while the accuracy of most of the algorithms is almost indistinguishable, there are substantial differences in their computational demands, ranging theoretically from polynomial to exponential complexity, and leading to substantial differences in computation time in realistic scenarios depending on the density and size of networks. Based on analysis of the algorithms and numerical simulations, we propose a hybrid approach providing competitive accuracy with improved computational efficiency. 
\end{abstract}

\maketitle

\begin{quotation}
Characterization of the structure of interactions in large heterogeneous systems based on observational data has become one of the dominant challenges across scientific fields. In many cases, measurements of the dynamical behavior is available, allowing inference of causal interactions among the subsystems by exploiting the principle of temporal precedence of the cause before the effect. Half a century ago Sir Clive Granger proposed a formal treatment of the problem of detecting such interactions, based on statistical testing of the improvement of the prediction of a target variable by a candidate source variable. This process has been generalized to nonlinear processes using the framework of information theory. However, the practical applicability of this methodology has been hindered by the need to properly account for all other potential intervening variables in the system, bringing in both computational and accuracy issues growing with network size. In this work we compare several prominent algorithms proposed recently for estimating causal structure in large networks. We introduce the algorithms within a common framework highlighting the similarities and differences, and compare their accuracy and computational demands on both simulated random networks and realistic examples derived from real-world brain and climate dynamics datasets. Finally, we suggest an algorithm with competitive accuracy and faster performance. 
\end{quotation}

\section{Introduction}
The study of complex dynamical systems is a growing area of research with applications in multiple fields ranging from neuroscience through genetics, ecology, social anthropology, informatics, economy and energetics to climate research -- see~\cite{Boccaletti2006} for an authoritative review including a range of application fields. 
This growth is fed by the increasing availability of large datasets with observational data from multiple subsystems of the studied systems, as well as by the rapidly increasing computational power of modern computers and progress in the algorithms for complex data analysis. 
A key principle in complex network research is viewing the system at hand as a network of interacting subsystems, with one of the central questions being that of estimating the pattern of mutual interactions of these. Notably, there is an ongoing transition from the previously prevailing study of purely statistical dependences between the subsystems (commonly denoted by the term 'functional connectivity', borrowed from the neurosciences~\cite{Friston1994}) to the quest of characterizing the pattern of the direct causal connections between the subsystems ('effective connectivity').

Note that for many complex systems, our knowledge of its structure and dynamics, although increasing at a tremendous pace, is very far from perfect. Therefore, the structure of interactions needs to be commonly estimated directly from observed time series. For instance in the case of the human brain, resolving the pattern of anatomical connections (structural connectivity) is still posing serious challenges, in particular using non-invasive methods~\cite{Schilling2018}. Moreover, depending on system parameters, the same structural connectivity can give rise to vastly different patterns of dynamical interactions~\cite{Hlinka2012EJN}. The pattern of functional connectivity can even change dynamically, giving rise to a progression of brain states, the detectability of which poses further methodological challenges~\cite{Hlinka2015, Preti2017}. In this context, the development and proper validation of methods for estimating the structure of causal interactions from observed time series is of key importance.

For stochastic processes, the problem of causal interaction discovery has been considered already by Norbert Wiener~\cite{Wiener1956} and later formulated by Sir Clive Granger in his famous concept of (Granger) causality. This generally states that a variable is to be considered causal with respect to some target variable, if its inclusion in a model improves the prediction of the target~\cite{Granger1969}. This operationalization has of course many practical and philosophical limitations, however has become commonly used at least as an interim approach in situations where testing causality by e.g. direct experimental manipulation is not readily available. For a more general discussion of causality and its inference we refer the reader to~\cite{Spirtes2000}. In principle, two key challenges appear in causal interaction discovery: the first is, under what conditions can the 'true' causality be uncovered from observational time series (such as observing all intervening variables), while the second lies in finding efficient algorithms for inference of the network of causal relations from data (particularly in the case of many dependent variables, short samples and potentially nonlinear relations). Thus, the practical application of Granger/Wiener conceptual solution of the former requires an effective algorithm for establishing causality from the observed time series of finite size to solve the latter challenge. Building on the original Granger's approach based on linear vector autoregressive processes, there has been a long line of attempts to widen the applicability of the principle by devising algorithms that would perform well also for nonlinear processes as well as in the situation of relatively large system size. 

The former problem of nonlinearity is commonly addressed by utilizing entropy as a general measure of uncertainty, or equivalently using mutual information as a measure of statistical dependence. This has motivated the definition of transfer entropy~\cite{Schreiber2000}, a special case of conditional mutual information~\cite{Palus2001} as a measure of causality in nonlinear dynamical systems. Similarly as in Granger causality, to avoid spurious inference due to indirect causation, the method can be extended by taking into account other potentially intervening variables by conditioning on other variables. 

However, the use of information-theoretical functionals only escalates the latter problem of dealing with high-dimensional data. Indeed, the complexity of standard binning algorithms grows exponentially with the dimension of the variables considered (as the number of multi-dimensional bins, at which the probability density is estimated, scales exponentially with the dimension). Alternative algorithms to estimate conditional mutual information based onnon-discretizing approaches such as kernel methods or k-nearest-neighbor-based estimators~\cite{Kraskov2004} also exist; however estimating entropy functionals is still a difficult task. To remedy this problem, several researchers have recently turned to schemes that reduce the number of conditions considered by some principled variable selection procedure. In particular, four such algorithms~\cite{Runge2012PRL, Kugiumtzis2013, Sun2015, Runge2017} were formulated in a way either directly inspired or at least resembling in some aspects the PC-algorithm~\cite{Spirtes2000}, in particular the first phase of PC, the skeleton discovery phase. 

These methods aim to construct a (directed) network representation of the systems causal structure; by evaluating the (conditional) mutual information from potential source variables to each target variable. 
Importantly, this network representation provides only a simplified picture of the full causal structure in a generic case, due to existence of higher-order (sometimes called polyadic, in contrast to dyadic) dependences.
Such potential approximation has been recently criticized, in particular concerning the problems of interpreting the reconstructed networks as information flows~\cite{James2016}.

However, proper theoretical treatment of higher order dependences as well as methods for their quantification from finite size samples are a matter of ongoing research, see~\cite{James2017, Martin2017, Allen2017} and references therein. While we believe, that building such theoretical fundamentals is key for proper interpretation of complex causal structures including higher-order dependences, in many practical situations such higher-order dependences may be negligible, particularly given their problematic estimability from small data samples. Therefore we believe that algorithms for construction of directed network representation of system causal structure will continue to be widely applied in practice, albeit this should be done with due caution.
While all the four above mentioned methods have been reported as reasonably performing on both simulated examples and real-world data, to the best of our knowledge, there has been no systematic theoretical and numerical comparison available that would help making informed choice concerning which of these algorithms to use in practical situations. However, we direct the attention of the readers at least to two recent works published during revisions of this manuscript, which contain in particular comparison of the naive fully multivariate TE with Sun's optimal causation entropy and the PCMCI~\cite{Runge2018a}, discussing several advantages of the PCMCI algorithm, and a recent review of causal inference with emphasis on applications in Earth sciences~\cite{Runge2019}.   

In this paper, we therefore set out to fill the gap by providing a comparison of structure, performance and computational demands of the reviewed algorithms, to help making informed choice in practice as well as to assess space and possible directions for improvement. Based on this comparison, we also propose a hybrid method that outperforms other methods in some scenarios in terms of computational demands under conserved accuracy.
The paper is structured as follows: after this introductory section, in section~\ref{sec:methods} we describe the compared methods as well as the selected procedure for statistical testing. Section~\ref{sec:data} describes the data that we use for assessing the accuracy and computational demands of the approaches. The results are shown in section~\ref{sec:results}, and the paper is finalized by a detailed discussion in section~\ref{sec:discussion} and final conclusion (section~\ref{sec:conclusion}).

%
%
%

\section{Problem statement and methods}
\label{sec:methods}

As mentioned above, under some assumptions causality may be defined in terms of reducing uncertainty of the prediction or in other words the conditional dependence of the target and source variable. These notions are conveniently formalized in terms of information-theoretical functionals.  
For two discrete random variables $X, Y$ with sets of values $\Xi$ and $\Upsilon$, marginal probability distribution functions $p(x), p(y)$, and joint probability distribution function $p(x,y)$, the Shannon entropy $H(X)$ is defined as
\begin{align}
 H(X)=-\sum_{x\in\Xi} p(x)\log p(x),
\end{align}
and the joint entropy $H(X,Y)$ of $X$ and $Y$ as
\begin{align}
 H(X,Y)=-\sum_{x\in\Xi}\sum_{y\in\Upsilon} p(x,y)\log p(x,y).
\end{align}

The conditional entropy $H(X|Y)$ of $X$ given $Y$ is 
\begin{align}
 H(X|Y)=-\sum_{x\in\Xi}\sum_{y\in\Upsilon} p(x,y)\log p(x|y).
\end{align}
The amount of common information contained in the variables $X$ and $Y$ is quantified by the mutual information $I(X;Y)$ defined as
\begin{align}
 I(X;Y)=H(X)+H(Y)-H(X,Y).
\end{align}
The conditional mutual information (CMI) $I(X;Y|Z)$
 of the variables $X,Y$ given the variable $Z$ is given as
\begin{align}
 I(X;Y|Z)=H(X|Z)+H(Y|Z)-H(X,Y|Z).
\end{align} 

Entropy and mutual information are measured in bits if the base of the logarithms in their definitions is 2. It is straightforward to extend these definitions to more variables, and to continuous rather than discrete variables. In practice, estimation of information-theoretical functionals for continuous variables is often carried out through their discretization by binning procedures, or alternative non-discretizing approaches such as kernel methods or k-nearest-neighbor-based estimators~\cite{Kraskov2004}. Alternatively, when data are considered sufficiently close to Gaussianity, estimates of linear quantities can be used -- in particular Pearson's correlation coefficient in place of mutual information and partial correlation in place of CMI.  

Let us now consider a random process $\lbrace \mathbf{X}_{t}| t \in \mathbb{Z} \rbrace$, where $\mathbf{X}_{t}$ is (for all $t \in \mathbb{Z}$) a multivariate random variable $\mathbf{X}_{t}=\left(X_{t}^{1}, \ldots, X_{t}^{n} \right)^{\top}$, with the random variable  $X_{t}^{i}$ indicating the state of element $i$ at the time $t$. Next we define $\mathbf{X}_{t}^{-}=\left(\mathbf{X}_{t-1}, \ldots,\mathbf{X}_{t-\tau_{\max}} \right)$, $\tau_{\max} \in \left[1, +\infty \right),$ 
which expresses the previous states of the system, similarly for each element of the system $X_{t}^{i-}=\left(X_{t-1}^{i}, \ldots,  X_{t-\tau_{\max}}^{i}\right).$

A natural way to quantify the causal effect of the variable $X_{t-\tau}^{j}$ on the variable $X_{t}^{i}$ conditioned on all other elements of the system $\mathbf{X}_{t}$ is the calculation of the CMI $I\left(X_{t}^{i};X_{t-\tau}^{j}|\mathbf{X}_{t}^{-} \smallsetminus X_{t-\tau}^{j} \right)$~\cite{Runge2017, Runge2018a}. 
Indeed, following the Granger's/Wiener's idea, a variable $X_{t-\tau}^{j}$ is to be considered causal with respect to the variable $X_{t}^{i}$, if $I\left(X_{t}^{i};X_{t-\tau}^{j}|\mathbf{X}_{t}^{-} \smallsetminus X_{t-\tau}^{j} \right)>0$. The reviewed causal discovery algorithms are thus trying to estimate the set of such nodes for a given target node, called the causal parent set: 
\begin{equation}
{N}_{X_{t}^{i}}=\lbrace X_{t-\tau}^{j}|\ I\left(X_{t}^{i};X_{t-\tau}^{j}|\mathbf{X}_{t}^{-} \smallsetminus X_{t-\tau}^{j} \right)>0 \rbrace.
\end{equation}

However, the evaluation of this CMI may in practice be unfeasible due to the problems with estimating high-dimensional information functionals, including computational demands and common inaccuracy of estimates from short time series samples~\cite{Kraskov2004, Runge2012PRL,Hlinka2013}. Therefore the below outlined algorithms for reduction of the dimension of the conditioning variable were proposed.

While Kugiumtzis et al. gave their algorithm a name (PMIME - partial mutual information from mixed embedding), two other algorithms were not introduced with an explicit name -- for simplicity and ease of orientation, we will refer to them as Runge's and Sun's algorithm throughout the paper, although abbreviation based on names of all original coauthors or procedural description might be advocated. The newer variant of algorithm proposed by Runge et al.~\cite{Runge2017} is then denoted in line with the original paper as PCMCI.

\subsection{Runge's algorithm and PCMCI}
The first of the studied algorithms is the algorithm introduced by Runge et al.~\cite{Runge2012PRL} 
Throughout this article we denote $\hat{n}=\lbrace 1, \ldots, n \rbrace,$ similarly $\hat{\tau}_{\max}=\lbrace 1, \ldots, \tau_{\max} \rbrace.$
The first step of the algorithm is to compute the mutual information $I\left(X_{t}^{i};X_{t-\tau}^{j}\right)$ for all $j \in \hat{n}$ and $\tau \in \hat{\tau}_{\mathrm{\max}}.$ Elements $X_{t-\tau}^{j}$ which share non-zero mutual information with $X_{t}^{i}$ form the set of potential causal parents of $X_{t}^{i}$, which we denote
\begin{equation}
\tilde{N}_{X_{t}^{i}}=\lbrace X_{t-\tau}^{j}|\ I\left(X_{t}^{i};X_{t-\tau}^{j} \right)>0 \rbrace.
\end{equation}
This set contains true causal parents, but also indirectly associated elements that have non-zero mutual information with the element $ X_ {t}^{i}$ for example because they are both influenced by some other element.
\begin{algorithm}[H]
\caption{The first phase of Runge's algorithm}\label{Runge1}
\begin{algorithmic}[1]
\State $\tilde{N}_{X_{t}^{i}} \gets \emptyset$
\For{$j \in  \hat{n} $}
\For{$\tau \in \hat{\tau}_{\max} $}
\If{$I\left(X_{t}^{i};X_{t-\tau}^{j}\right) >0$}
\State $\tilde{N}_{X_{t}^{i}} \gets \tilde{N}_{X_{t}^{i}}\cup \lbrace X_{t-\tau}^{j} \rbrace$
\EndIf
\EndFor
\EndFor
\end{algorithmic}
\end{algorithm}
In the second (reduction) phase of the algorithm these indirect links are therefore excluded from the set $\tilde{N}_{X_{t}^{i}} $. The natural way of this reduction is to determine for each element $X_{t-\tau}^{j}$ of the set $\tilde{N}_{X_{t}^{i}} $ the CMI
\begin{align}
\label{RungeInf}
I\left(X_{t}^{i};X_{t-\tau}^{j}|\tilde{N}_{X_{t}^{i}}\smallsetminus \lbrace X_{t-\tau}^{j} \rbrace \right)
\end{align}
and in the case when this information is equal to zero, exclude $X_{t-\tau}^{j}$ from the set of potential causal parents $\tilde{N}_{X_{t}^{i}}.$
However the size of the set of potential causal parents may be large; generally it may include up to $n \times \tau_{max}$ elements and the practical calculation of CMI may fail due to availability of only a short sample of the time series or due to the computational demands. Therefore in the reduction phase of Runge's algorithm, instead of computing the single conditional information (\ref{RungeInf}), the mutual information $I\left(X_{t}^{i};X_{t-\tau}^{j}|\tilde{N}^{m,k}_{X_{t}^{i}}\right)$  is computed over subsets $\tilde{N}^{m,k}_{X_{t}^{i}}$ of the original set $\tilde{N}_{X_{t}^{i}}$, where $m$ is the size of the subset and $k$ is the index of the subset.

\begin{algorithm}[H]
\caption{The second phase of Runge's algorithm}\label{Runge2}
\begin{algorithmic}[1]
\State $m \gets m_0$
\While{$ m < |\tilde{N}_{X_{t}^{i}}|$}
\For{$ k \in { |\tilde{N}_{X_{t}^{i}}| \choose m}$}
\If{$\tilde{N}^{m,k}_{X_{t}^{i}} \subseteq \tilde{N}_{X_{t}^{i}}$}
\For{$ X_{t-\tau}^{j} \in \tilde{N}_{X_{t}^{i}} \smallsetminus \tilde{N}^{m,k}_{X_{t}^{i}}$}
\If{$I\left(X_{t}^{i};X_{t-\tau}^{j}|\tilde{N}^{m,k}_{X_{t}^{i}}\right)=0$}
\State $\tilde{N}_{X_{t}^{i}} \gets \tilde{N}_{X_{t}^{i}} \smallsetminus \lbrace X_{t-\tau}^{j} \rbrace$
\EndIf
\EndFor
\EndIf
\EndFor
\State $m \gets m+1$
\EndWhile
\State ${N}_{X_{t}^{i}} \gets \tilde{N}_{X_{t}^{i}} $
\end{algorithmic}
\end{algorithm}

In detail, the second (reduction) phase of Runge's algorithm proceeds as follows. In the outer loop the parameter $m,$ which denotes the number of conditions in CMI, iterates upward from a predefined value $m_{0}> 0$. In the middle loop the parameter $k$ iterates through all the different subsets of size $m$ of $\tilde{N}_{X_{t}^{i}}.$ If for any $m$ and any $k$ is the CMI $I\left(X_{t}^{i};X_{t-\tau}^{j}|\tilde{N}^{m,k}_{X_{t}^{i}}\right)$ equal to zero than the element $X_{t-\tau}^{j}$ is removed from the set $\tilde{N}_{X_{t}^{i}}.$ If the size of the new set $\tilde{N}_{X_{t}^{i}}$ is less or equal to $m,$ algorithm terminates. Otherwise we increase $m$ by one and the algorithm continues. 

The Runge's algorithm was further developed to a more computationally efficient version of the algorithm introduced under the name PCMCI~\cite{Runge2017}, where PC stands for the names of Peter Spirtes and Clark Glymour, the authors of the PC algorithm~\cite{Spirtes2000}, and MCI stands for Momentary Conditional Independence. The initial phase of this algorithm is similar as the original algorithm (see algorithm~\ref{PC}), however the number of considered subsets $\tilde{N}^{m,k}_{X_{t}^{i}}$ is reduced. For a given $X_t^i$ and $X_{t-\tau}^j$ and for a given cardinality $m$ of subsets $\tilde{N}^{m,k}_{X_{t}^{i}}$, only $q_\text{max}$ CMIs in the form $I\left(X_{t}^{i};X_{t-\tau}^{j}|\tilde{N}^{m,k}_{X_{t}^{i}}\right)$ are assessed (instead of up to all possible $n\choose m$ combinations). Thanks to this reduction, the complexity of this algorithm changes from exponential to polynomial -- similarly to Sun's algorithm. The input parameter $q_\text{max}$ is selected by the user, for the current paper we used the setting $q_\text{max}=1$ that was applied for simulations in the original work~\cite{Runge2017}. Note that preferable choice may depend on the relative weight of required speed in the PC phase and the size of condition set entering the MCI phase. A major other difference is order-independence, which is achieved by not removing an independent parent immediately, but only after the loop (see line 16 in Algorithm 3). This leads in some situations to different results than for other common heuristi used in the classical PC-algorithm, where the variables 'conditioned out' are removed straight away, and therefore the result is not invariant with respect to the order of testing. Such approach is e.g. used in the later introduced FACDA algorithm, making it from this perspective order-dependent.

The last step of the PCMCI algorithm is the MCI step (see algorithm~\ref{MCI}). In this step, all elements $X_{t-\tau}^{j}$ (including those which were excluded in the PC phase) are tested against the output set of candidate variables from the PC phase. Moreover, the conditioning set in this phase does not contain only potential parents of the target $X_{t}^{i}$, but also potential parents of the source $X_{t-\tau}^{j}$, although only the $p_X$ strongest parents of the source are included, to limit the size of the condition. In this phase, for each element $X_{t-\tau}^{j},$ the CMI $I\left( X_t^{i}; X_{t-\tau}^{j}|{\tilde{N}}_{X_{t}^{i}} \cup \tilde{N}_{X_{t-\tau}^{j}}^{p_X} \smallsetminus \lbrace X^j_{t-\tau} \rbrace \right)$ is thus assessed.
A set of all $X_{t-\tau}^{j}$ for which is this CMI nonzero is declared as the set of causal parents of $X_{t}^{i}.$ The testing in the MCI phase is aimed to control the false positives rate at a predefined level; due to the inclusiong of the parents of the source and target, the tests should be valid even for highly autocorrelated variables, as effectively due to the conditioning only the relation between the residuals stripped of the autocorrelation is tested~\cite{Runge2017}.

In the PC phase, the authors recommend setting a relatively high value of parameter $\alpha$ which denotes level of statistical significance for which the $ H_{0} $ hypothesis (CMI is equal to zero) is rejected; in particular numerical examples show that  $\alpha>0.1$ leads to the false positive rate stabilizing around the expected level, while small $\alpha$ leading to too high false positive rate. In our simulations we use the setting $\alpha=0.2$. 

Further, in the final MCI phase, the use of false discovery rate (FDR) control~\cite{Benjamini1995} was recommended as a correction for multiple testing comparison. For a predefined FDR level, this effectively corresponds to using a corrected threshold that depends on the observed p-values across all the tests. To keep comparability with other methods, we use the range of the parameter $\theta \in \lbrace 0.1\%, \ldots,2.5\% \rbrace$ equally to other algorithms. Further, because of the potential problem with high dimensionality, the authors recommend to restrict the number of conditions $\tilde{N}_{X_{t-\tau}^{j}}$ with a free parameter $p_X$. We use the default setting $p_X=1$ recommended in similar simulations in the original study; i.e. we consider only one element of $\tilde{N}_{X_{t-\tau}^{j}}$, and for comparison a minimal choice $p_X=0$. In fact, choice of higher values had a detrimental effect on the accuracy of the algorithm in our simulations, see Figure~\ref{PCMCIPX}, we believe this is a design-choice of the author of the algorithm for a particular reason, namely achieving nominal FPR-control under autocorrelation. 





\begin{algorithm}[H]

\caption{Algorithm PCMCI - PC phase}\label{PC}

\begin{algorithmic}[1]
\State $\tilde{N}_{X_{t}^{i}}=\lbrace X^j_{t-\tau}| j \in \hat{n}; \tau \in \hat{\tau}_{\max} \rbrace$

\State $I^{\min}( X^j_{t-\tau})=+\infty \ \ \forall  X^j_{t-\tau} \in \tilde{N}_{X_{t}^{i}}$

\For{$m=0, \ldots, m_{\max}$}

\If{$|\tilde{N}_{X_{t}^{i}}|-1<m$}
\State Break for-loop
\EndIf
\For{$X^j_{t-\tau} \in \tilde{N}_{X_{t}^{i}}$}
\State $q=-1$
\ForAll{lexicographically chosen $S \subseteq \tilde{N}_{X_{t}^{i}} \smallsetminus \lbrace X_{t-\tau}^j \rbrace$ with $|S|=m$}
\State $q=q+1$
\If{$q \geq q_{\max}$}
\State Break from inner for-loop
\EndIf
\State$\left[I_{t-\tau}^j , p\text{-value} \right]=I\left(X_t^i,  X^j_{t-\tau}| S \right)$
\If{$|I|<I^{\min}\left(X^j_{t-\tau}\right)$}
\State$I^{\min}\left(X^j_{t-\tau}\right)=|I|$
\EndIf
\If{$p\text{-value}>\alpha$}
\State Mark $X^j_{t-\tau}$ for removal from $\tilde{N}_{X_{t}^{i}}$
\State Break from inner for-loop
\EndIf
\EndFor
\State Remove non-significant parents from $\tilde{N}_{X_{t}^{i}}$
\State Sort parents in $\tilde{N}_{X_{t}^{i}}$ by $I^{\min}\left(X^j_{t-\tau}\right)$ from largest to smallest
\EndFor

\EndFor

\end{algorithmic}

\end{algorithm}

\begin{algorithm}[H]

\caption{Algorithm PCMCI - MCI phase}\label{MCI}

\begin{algorithmic}[1]
\For{$X^j_{t-\tau} \in \lbrace X^j_{t-\tau}| j \in \hat{n}; \tau \in \hat{\tau}_{\max} \rbrace$}
\State{$\tilde{N}_{X_{t-\tau}^{j}}^{p_X} \gets$ first $p_{X}$ parents from $\tilde{N}_{X_{t}^{j}}$ shifted by $\tau$}
\If{$I\left( X_t^{i}; X_{t-\tau}^{j}|{\tilde{N}}_{X_{t}^{i}} \cup \tilde{N}_{X_{t-\tau}^{j}}^{p_X} \smallsetminus \lbrace X^j_{t-\tau} \rbrace \right)=0$}
\State{Mark $X^j_{t-\tau}$ for removal from $\tilde{N}_{X_{t}^{i}}$}
\EndIf
\EndFor
\State Remove non-significant parents from $\tilde{N}_{X_{t}^{i}}$
\end{algorithmic}

\end{algorithm}

\subsection{PMIME \& Sun's algorithm}
\label{ssec:PMIME}
Two other studied algorithms are the algorithm PMIME~\cite{Kugiumtzis2013} and Sun's algorithm~\cite{Sun2015}.
PMIME algorithm (partial mutual information from mixed embedding) was originally formulated in a more general setting for multiple time lags than the Sun's algorithm. However, in a basic setting (that means maximum time lag equal to 1 for every variable in the system) is this algorithm equivalent to the first phase of Sun's algorithm, which has been originally designed only for Markov processes of order one. 

\begin{algorithm}[H]
\caption{The first phase of Sun's algorithm}\label{Sun1}
\begin{algorithmic}[1]
\State $K \gets \lbrace X_{t-1}^j | \ j \in \hat{n}  \rbrace, \tilde{N}_{X_{t}^{i}} \gets \emptyset$, $I \gets +\infty$, $p \gets \emptyset$
\While{$I>0$}
\State $\tilde{N}_{X_{t}^{i}} \gets \tilde{N}_{X_{t}^{i}}\cup  p $
\For{$X_{t-1}^j \in \left( K \smallsetminus \tilde{N}_{X_{t}^{i}} \right)$}
\State $I_j \gets I\left( X_t^{i}; X_{t-1}^{j}|\tilde{N}_{X_{t}^{i}}\right)$
\EndFor
\State $I \gets \max \ I_j$
\State $\tilde{j} \gets \textbf{argmax} \ I_j$
\State $p \gets \lbrace X_{t-1}^{\tilde{j}} \rbrace $
\EndWhile
\end{algorithmic}
\end{algorithm}

This first phase of Sun's algorithm proceeds as follows.  
The initial step is to estimate the mutual information $I\left( X_t^{i}; X_{t-1}^{j} \right)$ for each element $j\in\hat{n}.$ If this mutual information is equal to zero for every $j\in\hat{n},$ the algorithm terminates. Otherwise the element with maximal mutual information is added to the (initially empty) set of potential causal parents $\tilde{N}_{X_{t}^{i}}.$  In the next steps, the CMI $I\left( X_t^{i}; X_{t-1}^{j}|\tilde{N}_{X_{t}^{i}}\right)$ is assessed for each $j \in \hat{n}$ for which $X_{t-1}^{j} \notin \tilde{N}_{X_{t}^{i}}.$ If this CMI is equal to zero for each $j,$ the algorithm terminates. Otherwise the element with maximal CMI is added to the set.

However, Sun et al. suggested (on the contrary to the authors of PMIME method) a necessity to include a second phase that would attempt to remove any spurious links, i.e. indirect links due to common mediator or false links due to common driver, included during the first phase. In the second (reduction) phase of Sun's algorithm, at each step $j$ the CMI $I\left(X_{t}^{i};X_{t-1}^{j}|\tilde{N}_{X_{t}^{i}}\right)$ is assessed. 
If this CMI is equal to zero, the element $X_{t-1}^{j}$ is excluded from the set of potential causal parents.
\begin{algorithm}[H]
\caption{The second phase of Sun's algorithm}\label{Sun2}
\begin{algorithmic}[1]
\For{$X^j_{t-1} \in \tilde{N}_{X_{t}^{i}}$}
\If{$I\left( X_t^{i}; X_{t-1}^{j}|{\tilde{N}}_{X_{t}^{i}} \smallsetminus \lbrace X^j_{t-1} \rbrace \right)=0$}
\State{$\tilde{N}_{X_{t}^{i}}=\tilde{N}_{X_{t}^{i}} \smallsetminus \lbrace X^j_{t-1} \rbrace$} 
\EndIf
\EndFor

\end{algorithmic}
\end{algorithm}
Similarly to Runge's original but unlike in the PCMCI algorithm, the order of testing of the elements from the set of potential causal parents $\tilde{N}_{X_{t}^{i}} $ may also influence the outcome of the Sun's algorithm.
In the original article~\cite{Sun2015} this fact is not discussed. In our implementation we use testing from the weakest element to the strongest. In this case we quantify the strength of the element $ X_{t-1}^{j} $ by the mutual information $I\left(X_{t}^{i};X_{t-1}^{j}\right).$ Note that the PCMCI is order-independent in that it avoids the need for order choice by only marking for removal instead of removing the explained parents straight away. In principle the PCMCI removes thus a superset of variables compared to removing directly during testing in any particular order; for more dicussion see~\cite{Runge2018a}.

\subsection{Relations between the algorithms}
\label{ssec:FACDA}
The description of the algorithms back to back already hints on their similarities and differences. In the following we shall make this comparison even more explicit and draw some suggestions and conclusions from this.

A naive approach to detecting the parent set of a given node would be to assess each potential parent node at a time by computing its information on the target node conditional on all other nodes. However, this would require computation of information functionals of high dimension, posing both computational and numerical problems. The reviewed algorithms sidestep this problem by limiting the candidate parent set in one way or another.

In particular, for each target node, all reviewed algorithms include an initial phase that generates a set of its candidate causal parents. This is done either at once by evaluating (unconditional) mutual information with the target (Runge's algorithm and PCMCI), or iteratively by evaluating the mutual information conditional on the already identified candidate parents (Sun's algorithm and PMIME). Then, a second phase may follow: potential candidates are removed by iterative testing of their added value (CMI) with respect to the rest of the candidate set (Sun) or with respect to its subsets of increasing size (Runge, PCMCI).

\begin{table}[]
\begin{tabular}{ l | l | l | l | l | l}
  Phase/algorithm  & PMIME    & Sun  & Runge & PCMCI & FACDA \\
         \hline
Forward  &  $\sim n^2$ &   $\sim n^2$  & $\sim n$      & $\sim n$      & $\sim n^2$      \\
Backward & $-$ &   $\sim n$  & $\sim 2^n$      &  $\sim n^2$     &  $\sim n$     \\
Repair   & $-$ &   $-$  &  $-$     & $\sim n$       &  $-$  \\  
\hline
Total  &  $\sim n^2$ &   $\sim n^2$  & $\sim 2^n$      & $\sim n^2$      & $\sim n^2$ 
\end{tabular}
\caption{Asymptotic worst-case number of CMI evaluations for obtaining the parent set of one node.}
\label{table:complexity}
\end{table}

The approach of Runge's algorithm is to first obtain a superset of the true parents by assessing the mutual information of each node with the target, and in the second phase iteratively try to remove them by conditioning on increasing subsets of other strong candidates. On the contrary to Runge's algorithm, in the first phase of Sun's algorithm the candidate parents are added one by one (i.e. evaluation of {\it conditional} mutual information (conditioned by elements of the current set of potential causal parents $\tilde{N}_{X_{t}^{i}}$) is used), and therefore after the first phase of Sun's algorithm the set of potential causal parents   $\tilde{N}_{X_{t}^{i}}$ should contain fewer (if any) indirect connected elements than after the first simple phase of Runge's algorithm, allowing to assess the fully conditioned mutual information. The two approaches thus principally differ in which phase they treat iteratively - the forward inclusion phase of the backward removal phase.
While the number of iterations is generally larger in Runge's algorithm, the number of evaluated nodes in each step of the iteraction is larger in the (iterative) first phase of Sun's algorithm; therefore it depends on the circumstances, which algorithm leads to less CMI evaluations in total.

In general, we expect Sun's algorithm to be more effective than Runge's for large dense networks due to its only polynomial complexity in network size. In particular, in the case when the $i$-th element of the system is influenced by all other elements, at maximum $n(n-1)/2 \sim n^2$ CMIs are evaluated in the first (more computational demanding) phase of Sun's algorithm. On the other side, in such extreme case, Runge's algorithm would pass (in the second phase) through all subsets of the (full) set of potential causal parents, in an attempt to 'condition out' the effect of a given candidate causal parent. In a system of $n$ elements this leads to assessing up to $2^{n-1}$ subsets;  leading to the worst case complexity exponential in $n$. 

Importantly, the PCMCI variant of Runge's algorithm largely remedies this weakness by limiting for each tentative parent the number of subsets of size $m$ it is tested against from above by a constant $q_{max}$, effectively providing a polynomial (quadratic) computational complexity of $\sim n^2 q_{max}$. The last phase added in the PCMCI algorithm to provide control of false positives at a predefined rate does not substantially affect the computation time. Note that (similarly as in the original Runge's algorithm), the algorithm could be further speeded up by limiting the maximum size of the condition 
by a constant $m_{max}$, leading to further potential speedup in exchange for higher false positive rate at the backward stage. 


\subsection{FACDA}
\label{ssec:}
Based on the theoretical analysis above, we conjecture that a key challenge for practically applicable algorithms is being able to deal with large dense networks. For this purpose, limiting oneself in each step to testing using only few strongest candidates instead of carrying out full search through conditioning sets might be a suitable heuristic. We implement this idea in a hybrid algorithm between the Runge's and PMIME algorithms, proposing thus a new Fast Approximate Causal Discovery Algorithm (FACDA), described in pseudo-code below.

\begin{algorithm}[H]

\caption{The first phase of FACDA algorithm}\label{FACDA}

\begin{algorithmic}[1]

\State  $K \gets \lbrace X_{t-\tau}^j | \ j \in \hat{n}; \ 0< \tau \leq \tau_{\max}  \rbrace$, $\tilde{N}_{X_{t}^{i}} \gets \emptyset$, $p \gets \emptyset$

\While{$\left( K \smallsetminus \tilde{N}_{X_{t}^{i}} \right) \neq \emptyset$}

\State $\tilde{N}_{X_{t}^{i}} \gets \tilde{N}_{X_{t}^{i}}\cup  p $

\For{$X_{t-\tau}^j \in \left(K \smallsetminus \tilde{N}_{X_{t}^{i}} \right)$}

\State $I_{t-\tau}^j \gets I\left( X_t^{i}; X_{t-\tau}^{j}|{\tilde{N}}_{X_{t}^{i}}\right)$

\If{$I^j_{t-\tau} = 0$}

\State $K \gets K \smallsetminus \lbrace X_{t-\tau}^{j} \rbrace$

\EndIf

\EndFor

\State $\left[j_m \tau_m \right] \gets \textbf{argmax} \  I^{j}_{t-\tau}$

\State $p \gets \lbrace X_{t-\tau_m}^{j_m} \rbrace $

\EndWhile

\end{algorithmic}

\end{algorithm}

\begin{algorithm}[H]

\caption{The second phase of FACDA algorithm}\label{FACDA2}

\begin{algorithmic}[1]
\For{$X^j_{t-\tau} \in \tilde{N}_{X_{t}^{i}}$}
\If{$I\left( X_t^{i}; X_{t-\tau}^{j}|{\tilde{N}}_{X_{t}^{i}} \smallsetminus \lbrace X^j_{t-\tau} \rbrace \right)=0$}
\State{$\tilde{N}_{X_{t}^{i}}=\tilde{N}_{X_{t}^{i}} \smallsetminus \lbrace X^j_{t-\tau} \rbrace$} 
\EndIf
\EndFor
\end{algorithmic}

\end{algorithm}

To understand the relation of FACDA to the algorithms presented earlier it is useful to introduce some concepts concerning feature selection procedures, in particular the forward selection, backward selection and early dropping. The former two denote commonly used heuristic algorithms, which are specific instances of stepwise methods. In the basic forward selection algorithm, the predictor/feature set is initiated as empty and in each step, the variable with maximal improvement in model fit is added to the set. The usual stopping criterion is lack of improvement in model fit by any of the remaining variables. Conversely, the backward selection algorithm initiates the feature set by the whole set of available features/variables, and iteratively removes the least relevant one. Combination of these basic heuristic approaches gives rise to a rich family of feature selection methods. For a simplified overview of the phases and computational complexities of the compared algorithms see Table~\ref{table:complexity}.

From this perspective, the first phase of the Sun's algorithm is a forward selection, while the second phase is a backward selection. Similarly, Runge's original algorithm consists of initialization of the feature set by a filtering step, with subsequent variant of backward selection (using iteratively increasing subsets, allowing potentially avoiding getting stuck in a local minimum). It is known that the forward selection may suffer from high count of false positives and relatively high computational demands for large data~\cite{Borboudakis2017}. These problems can be alleviated by narrowing down the search by filtering out variables that are deemed conditionally independent of the target given the current set of selected variables -- a heuristic recently introduced under the name Early Dropping~\cite{Borboudakis2017}. 

In this context, the first phase of FACDA algorithm entails a forward selection accelerated by applying the early dropping heuristic followed by the backward selection in the second phase of FACDA algorithm. For a more detailed review of iterative feature selection procedures in a general context not specific to causal network inference, we refer to the latter paper, that explicitly introduces and studies the properties of Forward-Backward selection with early dropping (FBED$^K$), an algorithm combining $K+1$ runs of the forward selection with early dropping with a final backward selection phase. 
Note that FBED$^1$ can be proven to correctly identify  the Markov blanket of the target variable under the faithfulness assumption and perfect statistical inference (for details and proof see~\cite{Borboudakis2017}). 
In this general nomenclature, FACDA would correspond to FBED$^\infty$ (or maybe FBED$^{ntau}_{max}$, as only finite set of possible parents is considered).  

In the case of causal network inference from time series, we are interested in whether the algorithms correctly detects the causal parent set for each node. Below we show a sketch of a proof of such convergence for the FACDA algorithm; note that similar arguments apply to other presented algorithms (apart from PMIME, which due to the lack of the second phase should provide a \emph {superset} of the causal parents under the below assumptions). We assume the following conditions:  causal sufficiency that assumes that common causes of all variables are measured, faithfulness, which ensures that the true parent $X_{t-\tau}^j \in {N}_{X_{t}^{i}}$ will not be eliminated by any set of other elements $S$, i.e.
$I(X_t^i; X_{t-\tau}^{j}|S)\neq 0,$ causal Markov condition, which guarantees that all elements $X_{t-\tau}^j\notin {N}_{X_{t}^{i}}$ will be eliminated by set of all causal parents ${N}_{X_{t}^{i}},$ i.e.
$I(X_t^i; X_{t-\tau}^{j}|{N}_{X_{t}^{i}} \smallsetminus \lbrace X_{t-\tau}^j \rbrace)=0,$
and perfect statistical inference. Of course, for finite size samples, statistical inference is imperfect, and therefore the prove below holds only asymptotically. For the exact definition of faithfulness and causal Markov condition see~\cite{Spirtes2000}.

First, we show that the first phase of the FACDA algorithm finds a superset of true causal parents  i.e. ${N}_{X_{t}^{i}} \subseteq \tilde{N}_{X_{t}^{i}}.$ Let us suppose that $X^j_{t-\tau}$ is the true parent of $X^i_t,$ i.e. $X^j_{t-\tau} \in  {N}_{X_{t}^{i}},$ but $X^j_{t-\tau} \notin  \tilde{N}_{X_{t}^{i}}.$ Hence there is a set $L\subset \tilde{N}_{X_{t}^{i}}$ such that $I(X_t^i; X_{t-\tau}^{j}|L)=0,$ which is in contradiction with faithfulness; hence $X^j_{t-\tau} \in  \tilde{N}_{X_{t}^{i}}.$

In the second phase, faithfulness guarantees that no true causal parent $X_{t-\tau}^j \in {N}_{X_{t}^{i}}$ will be excluded: let us assume that $X_{t-\tau}^j \in (\tilde{N}_{X_{t}^{i}}\smallsetminus {N}_{X_{t}^{i}})=M$; we will prove that $X_{t-\tau}^j$ will be eliminated in the second phase of FACDA algorithm. 
Based on step 2, element $X_{t-\tau}^j$ will be eliminated if $I(X_t^i; X_{t-\tau}^{j}|\tilde{N}_{X_{t}^{i}}\smallsetminus \{ X^j_{t-\tau}\})=0.$ From the first part we know that ${N}_{X_{t}^{i}} \subseteq \tilde{N}_{X_{t}^{i}},$ hence $\tilde{N}_{X_{t}^{i}} = {N}_{X_{t}^{i}}\cup M ,$ now from the causal Markov condition ensues that $I(X_t^i; M|{N}_{X_{t}^{i}})=0,$ because all elements in $M$ are not causal parents of $X_t^i,$ hence from the weak union $I(X_t^i; X_{t-\tau}^{j}|{N}_{X_{t}^{i}} \cup M \smallsetminus \lbrace X_{t-\tau}^j \rbrace)=0$ thus $I(X_t^i; X_{t-\tau}^{j}|{N}_{X_{t}^{i}} \smallsetminus \{ X_{t-\tau}^j \})=0$ and $X_{t-\tau}^j$ will be eliminated and then $\tilde{N}_{X_{t}^{i}}={N}_{X_{t}^{i}}.$


\subsection{CMI estimation and statistical testing}
In all presented algorithms it has to be repeatedly decided whether CMI is equal to zero or not. However such estimate from finite sample is generally nonzero even for independent variables, therefore a statistical test is required of the null hypothesis $H_{0}$ in the form
\begin{equation}
H_{0}:I \left( X;Y|Z \right)=0
\end{equation}
 at a predefined level of statistical significance $\theta.$

For speed and tractability reasons, in our numerical simulations we use only linear Gaussian models of random processes. Thus we can efficiently utilize an estimate of the CMI based on partial correlation $\rho\left( X,Y|Z \right)$:
\begin{equation}
\label{InfPart}
I \left( X;Y|Z \right)=-\frac{1}{2}\mathrm{log}\left( 1-\rho\left( X,Y|Z \right)^{2} \right)
\end{equation}
and thus we evaluate partial correlation instead of CMI. Note that in practice, the choice of estimator has substantial impact on computational complexity, see also the Discussion section.

To test if the CMI is zero, the authors recommend to use a permutation test, which does not assume normal distribution and independence of samples. In our data situation, due to the normality of the time series, we speed up the simulations by using the (approximate) default setting of the function \texttt{partialcorr} (MATLAB) in which the p-value is assessed by Student's t-test. Note that potential autocorrelation of the time series might lead to increased false positive rate in the individual tests.

\section{Data examples}
\label{sec:data}

The numerical comparison of the above presented algorithms is demonstrated on examples of vector autoregressive processes of order 1 (VAR(1) process) in the form
\begin{align}
\label{VAR(1)}
\mathbf{X}_{t}=\mathbb{A}\mathbf{X}_{t-1}+\mathcal{E}_{t}
\end{align}
where $\mathcal{E}_{t}$ denotes a white noise vector $\mathcal{E}_{t}=(\varepsilon^{1}_{t}, \ldots, \varepsilon^{n}_{t})^{\top}$ with covariance matrix $\text{Cov}(\mathcal{E}_{t})=\mathbb{I}.$ The structural matrix $\mathbb{A}$ carries information about the causal relationships. If we express the $i$-th row of this vector equation as
\begin{equation}
X_{t}^{i}=a_{i,1}X_{t-1}^{1}+ \ldots + a_{i,n}X_{t-1}^{n}+\varepsilon_{t}^{i},
\end{equation}
it is obvious that the $i$-th element of the system is affected by all elements for which $a_{i,j} \neq 0.$

In our numerical simulations we always work with a known matrix $\mathbb{A}$. From the expression (\ref{VAR(1)}) the time series of length $T$ are generated. These data serve us as the input for the studied algorithms whose output should ideally be the original matrix $\mathbb{A}$ or more precisely the binary structure of the matrix $\mathbb{A}$.

\subsection{Randomly connected networks}

We consider systems with random interaction structure $\mathbb{A}$ which we model by Erd\H{o}s-Rényi model of random graph (matrix). In this model the probability of presence of a direct link between each two elements is given by a predefined density value $D \in [0,1].$ Practically we fix the required density of the matrix $\mathbb{A}$ (percentage of the direct links) and assign a value of 1 to the corresponding number of randomly selected elements. This binary matrix is further normalized to ascertain stationarity of the process by multiplying it with a constant $\frac{s}{\lambda_{\max}},$ where $\lambda_{\max}$ is the largest eigenvalue (in absolute value) of the matrix $\mathbb{A}$, and $s\in(0,1)$ is an optional parameter. We set $s=0.8$ throughout the paper. 

\subsection{Realistically connected networks}
As real complex systems have structure that is neither random nor strictly regular, we further we use two datasets to provide realistic scenarios - one from the field of climatology and another from the field of neuroscience (described in detail bellow, more technical data description is available in a previous publication concerning small-world bias in correlation graphs of real-world networks~\cite{Hlinka2017Chaos}).
We make the approximation that these datasets correspond to  realizations of a VAR(1) process (\ref{VAR(1)}), therefore the elements of matrix $\mathbb{A}$ are estimated from the original data using linear regression. Then we retain in the matrix only a predefined percentage of the largest elements in the matrix, the rest is set to zero. This matrix is subsequently normalized by the constant $\frac{s}{\lambda_{\max}}$ and this matrix $\mathbb{A}$ defines the VAR(1) process (\ref{VAR(1)}). Note that both these datasets demonstrate also a high level of autocorrelation of the time series, an important property of real-world data that may affect the causal network recovery -- see Figures 2 and 4 in the Supplementary Material.

First we consider a 'climate network' constructed from regional daily time series. The network has 42 nodes and was obtained by thresholding the interaction matrix in a data-fitted VAR(1) model to $15$ percent density.
Details of the data origin and preprocessing are described in the Supplementary Material.

%

The second real-world example is a 'brain network'. The network has 90 nodes and was obtained by thresholding the interaction matrix in a data-fitted VAR(1) model to $5$ percent density.
We use data obtained as part of a study on healthy subjects brain activity.
The data describing the activity in 90 brain regions of 84 subjects were temporally concatenated in order to provide sufficiently long time series (20160 time points in total). Details of the data origin and preprocessing are described in the Supplementary Material.

Note that while in both the case of brain and climate, the approximation of the system by a linear vector autoregressive model of order one is clearly a daunting simplification of the original system, it has actually been previously shown to provide a surprisingly accurate representation of the observed system dynamics at commonly studied spatiotemporal scales~\cite{Hlinka2011Neuroimage, Hartman2011, Hlinka2014ClimDyn, Hlinka2015}.

\section{Numerical results}
\label{sec:results}
In this section we compare the studied algorithms using numerical simulations. In particular, we study the accuracy and the computational demands of the algorithms. Concerning the accuracy, as the algorithms are meant to estimate the binary structure of the matrix $\mathbb{A},$ we compare the ground-truth matrix $\chi_{0}({\mathbb{A}})$ defined as  
\begin{equation}
\left[ \chi_0 \left( \mathbb{A} \right)\right]_{i,j}= \left\{
\begin{array}{rl}
1 & a_{i,j}\neq 0 \\
0 & a_{i,j} =0 \\
\end{array}
\right.
\end{equation}
with the estimated matrix $\chi_{0}(\hat{{\mathbb{A}}}).$
The accuracy of each algorithm is described by two error measures: the false positive ratio $\varepsilon_+$ and the false negative ratio $\varepsilon_{-}$ given by
\begin{equation}
\varepsilon_{+} = \frac{ \#  \lbrace \left(i,j \right)| \chi_{0}\left(\mathbb{A} \right)_{i,j} =0 \wedge \chi_{0}(\hat{\mathbb{A}})_{i,j}=1  \rbrace }{\# \lbrace \left(i,j \right)| \chi_{0}\left(\mathbb{A} \right)_{i,j} =0\rbrace },
\end{equation}

\begin{equation}
\varepsilon_{-} = \frac{ \#  \lbrace \left(i,j \right)| \chi_{0}\left(\mathbb{A} \right)_{i,j} =1 \wedge \chi_{0}(\hat{\mathbb{A}})_{i,j}=0  \rbrace  }{\# \lbrace \left(i,j \right)| \chi_{0}\left(\mathbb{A} \right)_{i,j} =1\rbrace}.
\end{equation}

The computational demands are quantified by the total time of calculation.   The calculations were evaluated for a single core of the Intel(R) Xeon(R) CPU E5-2630 v2 2.60GHz processor; of course, mainly a relative interpretation of the computation time is informative, as the speed depends on many parameters of the hardware and implementation and in practice parallelization is easily available to speed up the computation. Therefore, we also provide the number of evaluations of CMI.

The numerical simulations proceeded as follows. According to the expression (\ref{VAR(1)}), the time series of length $T$ were generated. For robustness of evidence, 35 independent realizations of time series were generated for each specific parameter setting (35 random matrices were generated for the ER model). From these data the matrix $\chi_{0}(\hat{\mathbb{A}})$ was determined using each of the algorithms. We present the meadian values of the false positive ratio, false negative ratio and computational demands.

\subsection{White Noise \& Erd\H{o}s-Rényi model} 
The first studied model is a VAR(1) process with a random (Erd\H{o}s-Rényi model) structural matrix $\mathbb{A}=ER(n,D).$ Realization of this model is a binary matrix $\mathbb{A}$ of dimension $n \times n$ with a density $D$ of nonzero elements.  
For each of the 35 simulations, an independent realization of random matrix $\mathbb{A}$ was generated.

\begin{figure}
\includegraphics[width=0.9\columnwidth]{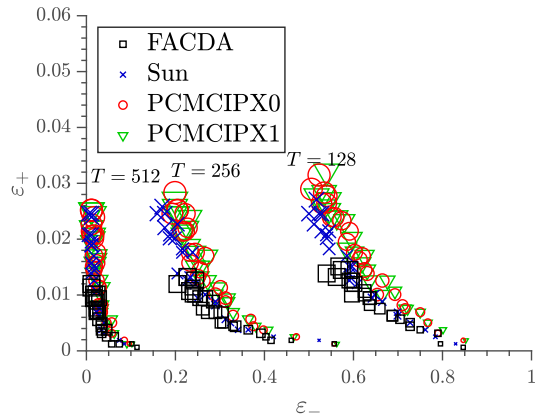}
\caption{Accuracy of the algorithms evaluated on simulations of VAR(1) process with random structural matrix $\mathbb{A}$ of density $D=10\%$ and size $n=42$.  Dependence of false positive ratio $\varepsilon_{+}$ on false negative ratio $\varepsilon_{-}$ for statistical test significance threshold $\theta \in \lbrace 0.1\%, \ldots,2.5\% \rbrace$ (denoted by increasing marker size) and time series length $T \in \lbrace 128, 256, 512\rbrace.$   }
\label{ER10}
\end{figure}

For the randomly connected VAR process, we choose a network size corresponding to the above described climate dataset ($n=42$), with density fixed to $D=10\%$. The corresponding Figure~\ref{ER10} shows the dependence of false positive ratio on false negative ratio for all algorithms. The simulation was carried out for a range of time series lengths: $T \in \lbrace 128, 256, 512\rbrace$ and a range of statistical threshold choices: $\theta \in \lbrace 0.1\%, \ldots,2.5\% \rbrace.$ Let us note that the accuracy for the PMIME algorithm corresponded almost perfectly to the results of Sun's algorithm and is thus not plotted separately. We also put into comparison only the last version of Runge's algorithm - PCMCI, due to its polynomial computational demands. The PCMCI algorithm is studied in two parameters settings which showed different accuracy. 
However, parameter setting in PCMCI does not have a significant effect on the computational demands, for this reason, we only include the results of numerical simulations of computational demands of version PCMCI ($p_x=0$).

As can be seen from Figure~\ref{ER10}, in line with reasonable expectations, the overall error of algorithms decreases with the increasing length of time series $T$. 
More interestingly, the accuracy of the algorithms seems to be comparable, 
only the PCMCI algorithm (for both parameter settings: $p_x=0$ and $p_x=1$) slightly differs from the others, this observation will be discussed in more detail in subsection~\ref{subsec:real} . The hypothetical curves of $\varepsilon_{+}$ as function of $\varepsilon_{-}$ largely overlap. However, for a fixed value of $\theta$, these algorithms are not comparable in their error rates  -- the algorithm of Sun/PMIME and PCMCI give more false positives and less false negatives, i.e. are more liberal. 
Conversely, our algorithm FACDA is more conservative. 

\begin{figure}
\includegraphics[width=0.9\columnwidth]{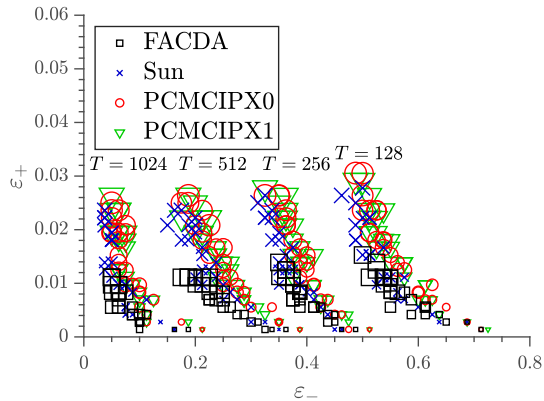}
\caption{Accuracy of the algorithms evaluated on simulations of a VAR(2) process with random structural matrices of density $D=10\%$ and size $n=20$. Visualization and settings as in Figure~\ref{ER10}.   }
\label{VAR2ER10}
\end{figure}

Similar result concerning accuracy is reproduced also for denser networks -- see Supplementary Material Figure 5 for results obtained for a corresponding simulation using a network density $D=15\%$. Indeed, here longer time series were needed to achieve comparable accuracy. Comparison of the algorithms was also carried out on an example of VAR(2) model (see Figure~\ref{VAR2ER10}) with both lag-1 and lag-2 matrices having density 10 percent and network size $n=20$ nodes. Stationarity of the corresponding VAR(2) process was again done by their scaling to assure that the leading eigenvalue of corresponding VAR(1) matrix is fixed to 0.8.

\begin{figure}
\includegraphics[width=0.9\columnwidth]{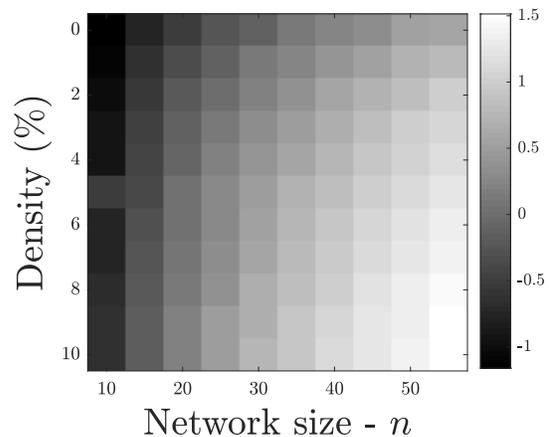}
\caption{Computational demands of network estimation by Sun's algorithm. Results for network sizes $n \in \lbrace 10,15,\ldots, 55\rbrace$ and densities $D\in \lbrace0, \ldots, 10\% \rbrace.$ Decadic logarithm of median computation time in seconds shown in grayscale. Time series of size $T=1024$ were generated from VAR(1) process with random structural matrix $\mathbb{A},$  statistical threshold set to $\theta=0.1$.}
\label{TS}
\end{figure}

Comparison of computational demands was carried out for network sizes $n \in \lbrace 10,15,\ldots, 55\rbrace $ and densities $D\in \lbrace0, \ldots, 10\% \rbrace.$ For density $D=0$ the VAR(1) process is equivalent to the vector form of a white noise process. As a baseline example, the total computation time for Sun's method is shown in Figure~\ref{TS}. In line with the theoretical expectation, the computational demands grow substantially with increasing network size and density. 


\begin{figure}
\includegraphics[width=0.9\columnwidth]{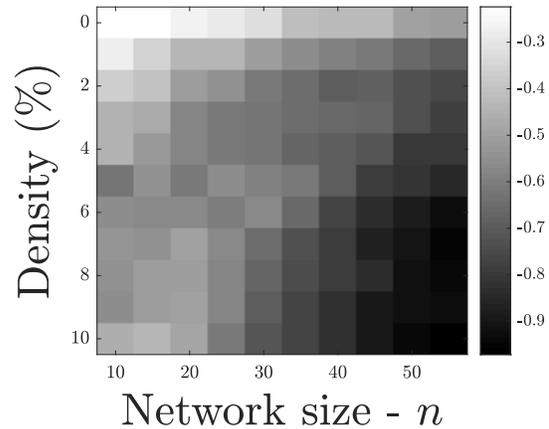}
\caption{Relative computational demands of network estimation by PCMCI with respect to Sun's algorithm. Visualization and parameter settings as in Figure~\ref{TS}. }
\label{RateTQ1Sun}
\end{figure}

\begin{figure}
\includegraphics[width=0.9\columnwidth]{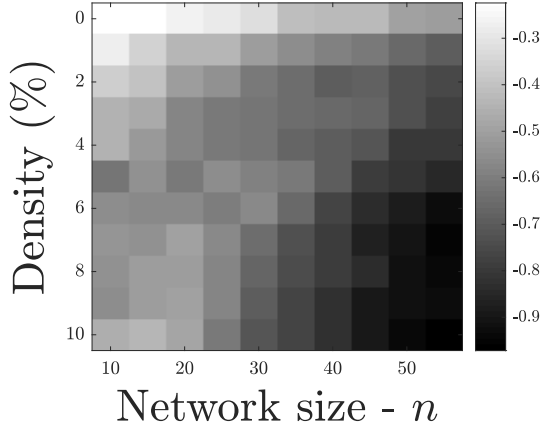}
\caption{Relative computational demands of network estimation by FACDA and Sun's algorithm. Visualization and settings as in Figure~\ref{TS}. }
\label{FACDASunT}
\end{figure}

As documented in Figure~\ref{RateTQ1Sun}, the new PCMCI algorithm by Runge et al. provides, particularly for the large dense networks, a substantial speedup against not only the original Sun's algorithm. Similar if not better performance as the PCMCI is provided by our algorithm FACDA. Detailed comparison with respect to the Sun's algorithm is shown in Figure~\ref{FACDASunT}.

\begin{figure}
\includegraphics[width=0.9\columnwidth]{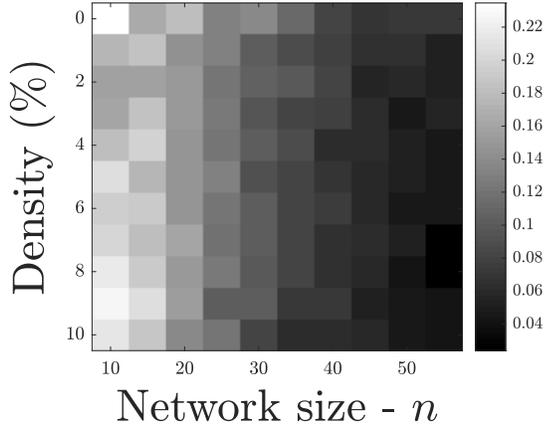}
\caption{Relative computational demands of network estimation by Sun's with respect to PMIME algorithm. Median computation time of Sun's algorithm divided by median computation time of PMIME is shown. Other visualization and parameter settings as in Figure~\ref{TS}.}
\label{TSP}
\end{figure}

As described in Subsection~\ref{ssec:PMIME}, in the current simplified setup, the PMIME algorithm is equivalent to the first phase of Sun's algorithm. For this reason, PMIME is necessarily less computationally demanding than Sun's algorithm. However, Figure~\ref{TSP} suggests that for large networks, this difference becomes negligible.


\subsection{Realistic datasets}
\label{subsec:real}
\begin{figure}
\includegraphics[width=0.9\columnwidth]{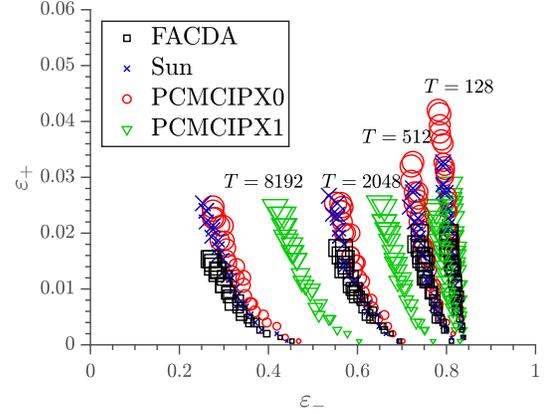}
\caption{Accuracy of the algorithms evaluated on VAR(1) process from climate data with structural matrix $\mathbb{A}$ of density $D=15\%.$ Dependence of false positive ratio $\varepsilon_{+}$ on false negative ratio $\varepsilon_{-}$ for parameter $\theta \in \lbrace 0.1\%, \ldots,2.5\% \rbrace$ and time series of length $T \in \lbrace 128, 512, 2048, 8192\rbrace.$}
\label{KL15}
\end{figure}

\begin{figure}
\includegraphics[width=0.9\columnwidth]{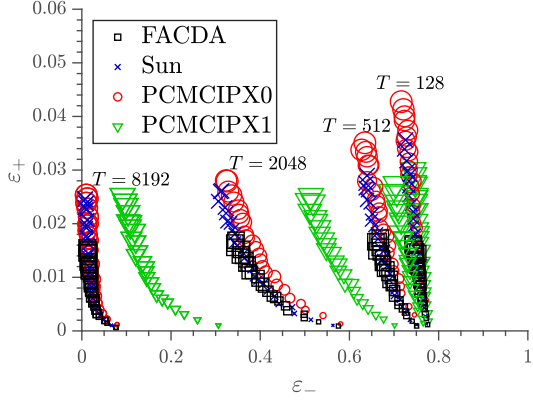}
\caption{Accuracy of the algorithms on VAR(1) process from brain data with structural matrix $\mathbb{A}$ of density $D=5\%.$ Visualization and settings as in Figure~\ref{KL15}.}
\label{BrainAcc}
\end{figure}

Further simulations were carried out with a structural matrix derived from realistic datasets including a climatic dataset ($n=42$) with density of structural matrix $D=15\%$ and a brain dataset ($n=90$) with density of structural matrix $D=5\%.$ The corresponding structural matrices are shown in the Supplementary material Figures 1-4. Numerical assessment of accuracy of the algorithms was carried out for parameter settings $T \in \lbrace 128, 512, 2048, 8192\rbrace$ and $\theta \in \lbrace 0.1\%, \ldots,2.5\% \rbrace.$  
The simulation results are shown in Figure~\ref{KL15} and Figure~\ref{BrainAcc} respectively. Similarly to the Erd\H{o}s-Rényi model, the simulations also suggest that PCMCI ($p_x=0$), Sun's and FACDA algorithms are comparable in their accuracy.
For both realistic datasets, the achieved accuracy was lower than for the randomly connected networks analyzed in the previous section (for a given time series length). This can be ascribed to the heterogeneous strength of links in realistic datasets, with a substantial proportion of relatively weak links, that are difficult to estimate correctly from short samples. 
As in the case of the Erd\H{o}s-Rényi model, we simulate also VAR(2) process modeling the 'climate network', further supporting the previous conclusions, see Supplementary Results Figure 7.

However, the PCMCI ($p_x=1$) differs from the other three in that while it  achieves lower false positive ratio (which is fixed on the value of theta as can be seen in figure \ref{BrainF}), this is more than outweighted by increases in false negative ratio. While this effect is present also in the simulated ER random networks in a weaker form, it is most clear for these inhomogeneous networks.

\begin{figure}
\includegraphics[width=0.9\columnwidth]{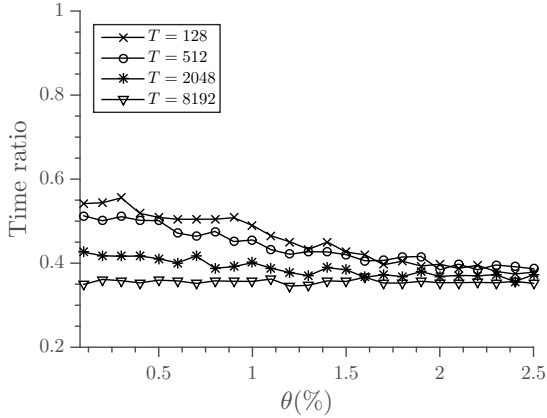}
\caption{Ratio of the total time of network estimation of PCMCI with respect to Sun's algorithm as function of statistical threshold $\theta.$ Results for time series of length $T \in \lbrace 128, 512, 2048, 8192 \rbrace$ generated from VAR(1) process with structural matrix $\mathbb{A}$ (density $D=15\%$) from the climate dataset.}
\label{TimeKL15PCMCI}
\end{figure}

The results of computational demands for the 'climate network' with $15$ percent density are shown in Figure~\ref{TimeKL15PCMCI}. 



\begin{figure}
\includegraphics[width=0.9\columnwidth]{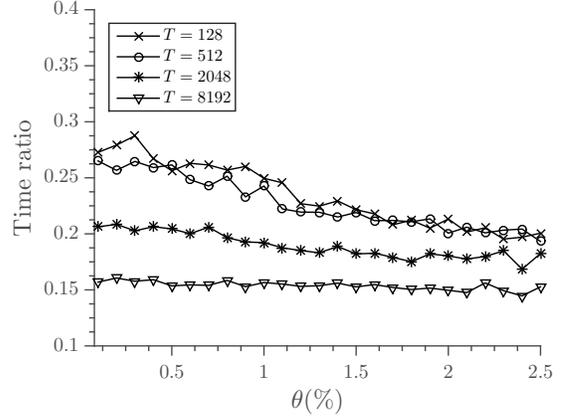}
\caption{Ratio of the total time of network estimation of FACDA with respect to Sun's algorithm as function of  statistical threshold $\theta.$ Visualization  as in Figure~\ref{TimeKL15PCMCI}.}
\label{TimeFACDASunKL15}
\end{figure}

\begin{figure}
\includegraphics[width=0.9\columnwidth]{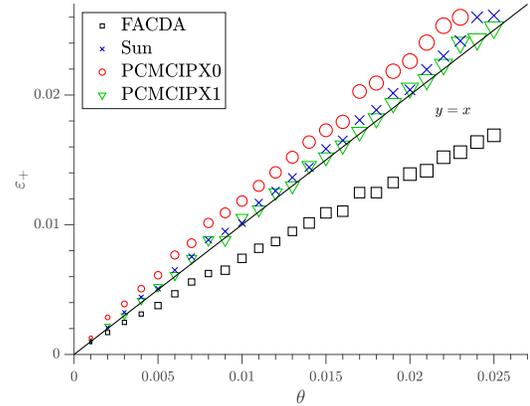}
\caption{Accuracy of algorithms inn simulations of VAR(1) process from brain data with structural matrix of density $D=5\%.$ Dependence of false positive ratio $\varepsilon_{+}$ on parameter $\theta$ for time series length $T=2048.$}
\label{BrainF}
\end{figure}

While the worst case complexity is polynomial for both (PCMCI and Sun's) algorithms, particularly for low values of $q_\text{max}$ the PCMCI is faster.
FACDA algorithm provides similar if not better performance as the PCMCI algorithm, that is substantial speedup particularly for large dense networks. Detailed comparison with respect to the Sun's algorithm is shown in Figure~\ref{TimeFACDASunKL15}. Qualitatively equivalent results were obtained for other settings, see results for the 'brain network' with density $5\%$ in the Supplementary Materials Figure 6.




\begin{figure}
\includegraphics[width=0.9\columnwidth]{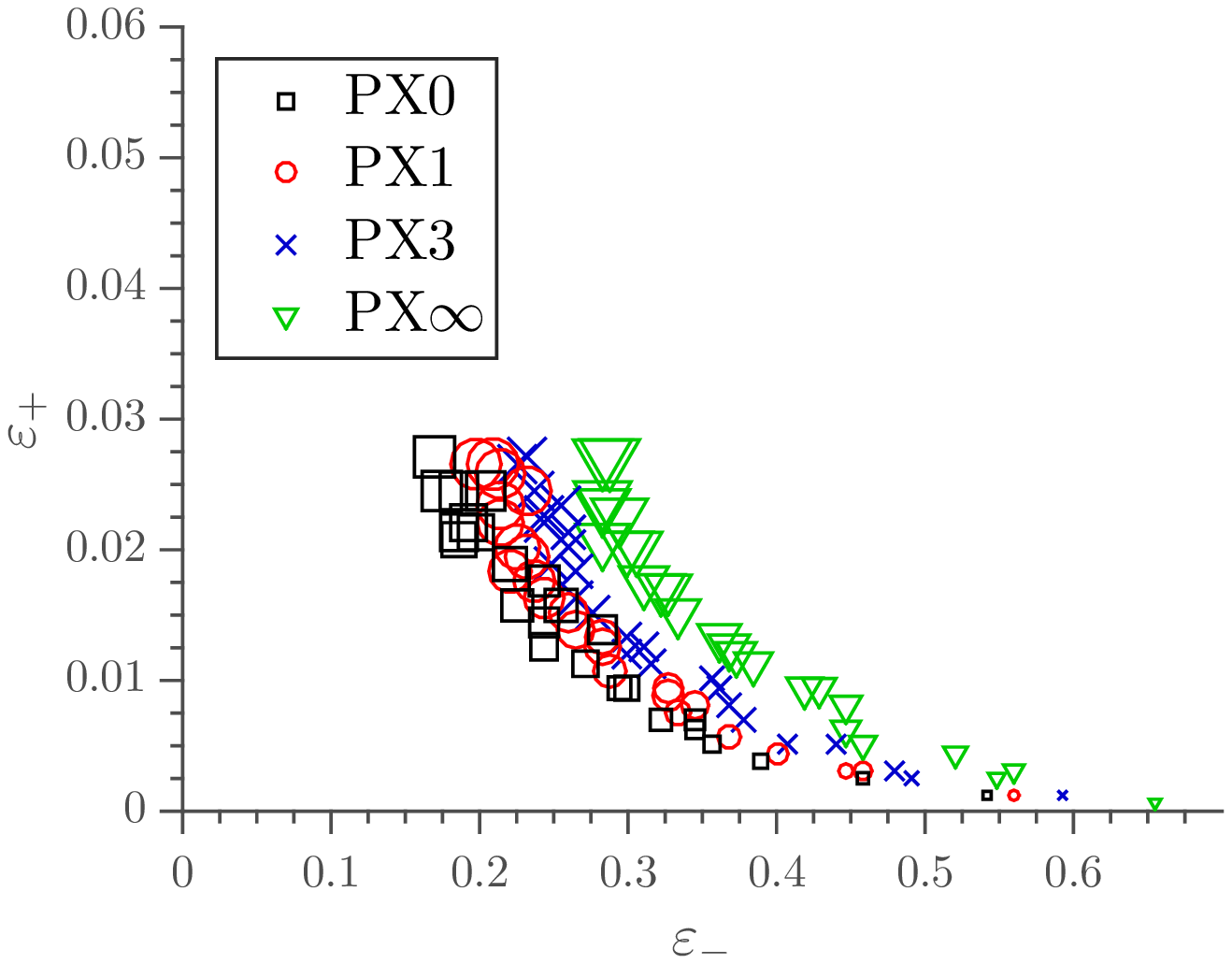}
\caption{Accuracy of the PCMCI algorithm in dependence on parameter $p_X$, evaluated on simulations of VAR(1) process with random structural matrix of density $D=10\%,$ size $n=42$, sample length $T=256.$}
\label{PCMCIPX}
\end{figure}

\section{Discussion}
\label{sec:discussion}
The comparison of the algorithms (PCMCI~\cite{Runge2017}, Sun's~\cite{Sun2015} and PMIME by Kugiumzis~\cite{Kugiumtzis2013}) has shown, that in realistic settings, they do not substantially differ in accuracy (PCMCI ($p_x=0$)), across a range of systems and parameter settings. 


The newly introduced FACDA method appears to keep the improved computational performance without the detrimental effects on the accuracy, giving similar or better results than the original three methods.

Notably, there are substantial differences in computational demands among the methods. 
Only a subtle difference is between the PMIME and Sun's method, given by PMIME missing a second phase -- for larger networks this difference appeared negligible. 

Runge's original algorithm is remedied in its new variant, PCMCI, that limits the number of tests in each cycle to $q_\text{max}$, leading thus to maximally polynomial complexity. Similar or even stronger improvement is also achieved in the FACDA approach, that provided here up to an order of magnitude speedup over Sun's algorithm in the case of the large dense networks. Notably, while the FACDA method can be considered as derived from the Sun's approach, a theoretical comparison shows that it is conceptually hybrid between this and PCMCI, being equivalent to PCMCI with several alterations: fixing $q_\text{max}=1$, accepting the strongest candidate in each cycle without testing, defining the strength in each cycle by the current CMI instead of the lowest value achieved so far, marking for removal instead of removing straight away (achieving thus order-independence, and therefore irrelevance of order of testing candidates in line 6 of the PC phase of PCMCI; while FACDA chooses lexicographic order and its change would generally alter the specific results) and omitting the final (MCI) phase. The specific or combined effect of these variations is a topic for further study that may lead to potential improvement of the algorithms. As it may depend on system parameters, one of the possible avenues is to provide adaptive data-informed algorithms.


A somewhat open problem is the choice and overall statistical interpretation of the threshold parameter controlling the leniency of the statistical test for the inclusion (or exclusion) of a candidate parent. Firstly, setting it to a given value $\theta$ does not guarantee fixing the resulting false positive rate to such value (not even asymptotically), due to the complex multiple testing procedure giving rise to the resulting networks -- unless a final 'repair' phase is included, as in the PCMCI algorithm. Secondly, as this bias differs between methods, setting the same $\theta$ leads to different behaviour of the methods, as they work at a different point along their receiver operating curve. In particular, for a fixed $\theta$, FACDA typically gave less false positives, but more false negatives, so the overall procedure can be considered as more conservative for a fixed $\theta$. However, similar performance can be obtained from Sun's method and PMIME by decreasing their $\theta$ parameter.  

The MCI phase of PCMCI algorithm guarantees (asymptotically) the control of false positive rate at the predefined level $\alpha$. However, it is likely responsible for the overall decreased performance (particularly because testing conditional independences is carried out with respect to parents of both source and target, therefore working with larger condition sets and smaller estimated effects). This is even stronger for high maximum included number of source parents $p_X$, as is shown in Figure~\ref{PCMCIPX}.

On the other side, apart from estimating causal strength stripped of the autocorrelation effects, the MCI phase of PCMCI has the advantage that the false positive rate is controlled asymptotically at the prescribed level given by the statistical threshold in this phase, see Figure~\ref{BrainF}.  This was in the simulations approximately true also for the Sun's method, while FACDA has lower-than-prescribed false positive rates, which can be attributed to the early reduction of the candidate set. Results for shorter time series are shown in the Supplementary Material Figure 9; note that for small sample the parametric partial correlation test may be imprecise and the use of some permutation scheme may be more suitable for exact control of false positive rate; for longer time series (Supplementary Material Figure 8) the Sun's and PCMCI methods false positive rates converge to the prescribed value.


Notably, the provided numerical comparisons were carried out using linear vector autoregressive processes. This is a standard type of stochastic system used in the original papers introducing the methods, as it allows more extensive numerical comparisons due to the possibility of very efficient estimation of CMI even in high dimensions through the use of partial correlation. Indeed, for Gaussian processes the transfer entropy is equivalent to Granger causality~\cite{Barnett2009}, which supports the use of linear methods for data that are deemed reasonably close to Gaussian; however even in the linear case, reduction of number of conditions may be computationally beneficial. 

When the assumption of Gaussianity is not suitable, other estimators of the (conditional) mutual information need to be used, and this may further (detrimentally) affect both accuracy and computational demands of the algorithms; in ways that would depend on the particular estimator in use. In this sense, our results provide only a rough guide, valid as long as this extra demands are comparable across methods. In~\cite{Runge2018a,Runge2018b} experiments with kNN estimators and also other versions find considerable trade-offs in runtime, showing that sometimes it's faster to run a full-conditioning, sometimes not, offering interesting insights while providing space for development of adaptive approaches.

Apart from the general argument mentioned above, it is important to note that the linear (Gaussian) approximation of the CMI by partial correlation is indeed commonly used in causal network discovery practice. This pragmatic choice is for many systems indeed substantiated by quantitative evidence concerning near-Gaussianity of the studied time series, e.g. for the brain activity data measured by functional magnetic resonance imaging~\cite{Hlinka2011Neuroimage, Hartman2011} or climate temperature time series~\cite{Hlinka2014ClimDyn, Hlinka2013}.

Notably, even in the linear setting, the compared algorithms become computationally intractable for networks larger than about a hundred of nodes (particularly for larger network densities). This constitutes a serious limitation. For such situations, some amendments to the methods or use or development of other algorithms would be necessary. 
As a sidenote, a more detailed analysis suggests, that 
the limiting factor is not necessarily the overall link density, but the maximum in-degree, i.e. the maximal (candidate) parent set. Conversely, further speedup can be of course achieved e.g. by limiting the maximum size of the conditioning set or number of tested conditions, with the trade-off of larger false positive rate in the forward phase. Such parametric variation of the algorithm (explicitly suggested e.g. in PCMCI) effectively renders a family of methods, of which we tested only some recommended default variants.


Alternatively, the use of some weak heuristic assumptions may allow effective estimation of even much larger causal networks. An example is the successful estimation of the global climate network of causal interactions based on temperature measurements in 2512 equidistantly distributed nodes on the globe, based on estimation of Granger causality and selection of the outgoing link with highest Granger causality index for each node~\cite{Hlinka2017Chaos2}. Despite being apparently simplistic, this approach was well suited to the data and allowed the discovery of a smooth causal flow in the global climate network, that until then went unnoticed due to the inability of general-purpose causal network analysis methods to deal with such a large network. A yet another alternative approach is to reduce the dimension before constructing the network by a suitable procedure~\cite{Vejmelka2015,Runge2015}.

Of course, while we have compared the most prominent algorithms introduced within the complex networks community, there are other alternatives for causal network estimation, building on the concepts of nonlinear extensions of Granger causality analysis~\cite{Marinazzo2008} as well as applying regularization procedures~\cite{Arnold2007}. From the breadth of reports concerning network reconstruction we direct the attention of the reader at least to a recent work including interesting combination of these ideas and also a useful comparison and overview of some of these alternative methods~\cite{Yang2017}. A yet another family of methods for detection of causal interactions has developed in the area deterministic nonlinear dynamics, we refer the reader to comparative reviews~\cite{Pereda2005, Chicharro2009} for detailed discussion of a range of methods formulated for detecting causality in the bivariate case. Generalization of many nonlinear methods to fully multivariate setting is not readily available and is a matter of further research, however, for some indices it is already available. Apart from the use of CMI in multivariate setting, and already mentioned nonlinear kernel Granger approaches, another recently proposed principled Granger causality generalization is the definition of nonlinear Granger causality through local linearization~\cite{Wahl2016}. This approach provides a consistent and well defined generalization of linear Granger causality and lends itself to straightforward generalization to conditional~\cite{Wahl2017} and multivariate setting. 

The research in causal network discovery is a very dynamic field that is being addressed by experts from multiple fields, sometimes not necessarily aware of the developments in other disciplines. We believe that further progress will be made by cross-fertilization between various approaches including the methods compared in this paper (PC-algorithm variants or other iterative approaches), regularization techniques and Bayesian inference with context-informed priors. 

On the other hand, the suitability or at least proper interpretation of the characterization of the causal structure by a (directed) network has been recently problematized, see e.g. Ref.~\cite{James2016}. In the current paper, we have on purpose used a system example that does not contain higher-order (polyadic) dependences; the true causal structure is thus unambiguous and well represented by a directed graph. However to at least comment on the potentially difficult to interpret behavior of the considered algorithms when applied to processes with higher-order dependences, we invite the reader to consider the case of a process given by $Y_{t+1}=XOR(X_t,Z_t)$, with $X_t, Z_t$ being independent boolean variables with $p(0)=p(1)=0.5$. There, no coupling would be detected by the presented algorithms, as they all start with assessing unconditional mutual informations $I(Y_{t+1},X_t)$ and $I(Y_{t+1},Z_t)$, both of which are equal to zero (of course, in practice, random sampling would give rise to some 'false' detections). While this example may seem singular or too artificial, less trivial and more realistic could be considered and we agree with~\cite{James2017} that for general complex systems, particularly with substantial higher-order interactions, we may need more fundamental theoretical formalisms as well as algorithms that would allow suitable representation of the causal structure going possibly beyond bivariate dependences -- we refer the reader to recent works in this area~\cite{James2017, Martin2017, Allen2017} and references therein for discussion of the possible avenues.


\section{Conclusions}
\label{sec:conclusion}

We have carried out a comparison of several prominent algorithms for causal network reconstruction. While they were originally introduced within slightly different contexts (such as explicit inclusion of arbitrary temporal lag or multivariate target variables), these algorithms share common ground and are related to the general PC-algorithm. The main difference between the algorithms is whether they use correlation or partial correlation for deciding on inclusion into the set of candidate parents in an initial phase of the algorithm, and whether they include a second phase for removal of indirect links from this set of candidate parents. By testing the algorithms using simulations of Gaussian processes on randomly and realistically connected networks (motivated by neuroscientific and climate data), we have shown that in practical usage these algorithms provide close to equivalent performance. However, the methods differ in their computational demands, most substantially for large networks: for sparse networks, selection of candidate parents by a single run of mutual information can be more effective; for denser networks, using CMI in the first phase provides substantial speedup through decreasing the size of the candidate parents set. However, similar computational demands can be achieved in a reduction phase by limiting the testing to a heuristically selected non-exhaustive sampling of the strongest conditions. We also commented on the problems of control of false positives and order-dependence, although for detailed discussion, we referred the reader to other works. 

Finally, we have proposed a new hybrid Fast Approximate Causal Discovery Algorithm (FACDA), designed for improved performance while essentially conserving accuracy.  Despite the current progress in algorithms, large and dense networks represent a  challenge for all presented methods, constituting a key open problem in causal network analysis.  

\section{Supplementary material}
See Supplementary material for the results of further analysis described in the Results section.

\begin{acknowledgments}
This work was supported by the Czech Health Research Council Projects No. NV15-29835A, No. NV15-33250A, and No. NV17-28427A; and by project Nr. LO1611 with a financial support from the MEYS under the NPU I program.
We thank Nikola Jajcay, David Hartman and David Tome\v{c}ek for valuable help with data preparation. 
\end{acknowledgments}

\bibliography{references2} 
\end{document}